\documentclass[preprint,preprintnumbers,nofootinbib]{revtex4-1}
\usepackage{amsmath} 
\usepackage{amsfonts} 
\usepackage{amssymb}
\usepackage{graphicx} 
\usepackage{color}
\usepackage{mathtools}
\usepackage{physics}

\newcommand{\be}{\begin{eqnarray}}
\newcommand{\ee}{\end{eqnarray}}

\newcommand{\eqn}[1]{\begin{equation} #1 \end{equation}}
\newcommand{\al}[1]{\begin{align} #1\end{align}}
\newcommand{\ri}{\mathrm{i}}
\newcommand{\Pii}{\Pi_i}
\newcommand{\Pione}{\Pi_1}
\newcommand{\Pitwo}{\Pi_2}

\newcommand{\Pir}{\Pi_\mathrm{r}}
\newcommand{\Pia}{\Pi_\mathrm{a}}
\newcommand{\si}{s_i}
\newcommand{\sone}{{s_1}}
\newcommand{\stwo}{{s_2}}
\newcommand{\sr}{s_\mathrm{r}}
\newcommand{\sa}{s_\mathrm{a}}
\newcommand{\up}{+}
\newcommand{\down}{-}
\newcommand{\updown}{\pm}
\newcommand{\GRAB}{M}

\begin{document}

\preprint{KEK-TH-2426}

\title{Complementarity and causal propagation of decoherence by measurement  in relativistic quantum field theories} 
\author{Yoshimasa Hidaka$^{a,b,c,d}$}
\email{hidaka@post.kek.jp}
\author{Satoshi Iso$^{a,b,e}$} 
\email{satoshi.iso@kek.jp}
\author{Kengo Shimada$^a$}
\email{skengo@post.kek.jp}
\affiliation{
$^a$ Theory Center,  High Energy Accelerator  Research Organization (KEK), Oho 1-1, Tsukuba, Ibaraki 305-0801, Japan \\
$^b$ Graduate University for Advanced Studies (SOKENDAI), Oho 1-1, Tsukuba, Ibaraki 305-0801, Japan\\
$^c$ Department of Physics, Faculty of Science, University of Tokyo, 7-3-1 Hongo Bunkyo-ku Tokyo 113-0033, Japan\\
$^d$ RIKEN iTHEMS, RIKEN, Wako 351-0198, Japan \\
$^e$ International Center for Quantum-field Measurement Systems for Studies of the Universe and Particles (QUP), KEK, Oho 1-1, Tsukuba, Ibaraki 305-0801, Japan
}
\begin{abstract} 
Entanglement generation by Newtonian gravitational potential between objects
has been widely discussed to reveal the quantum nature of gravity. 
In this paper, we perform a quantum field theoretical analysis of 
a slightly modified version of the gedanken experiment 
by Mari and co-workers [A. Mari {\it et al}., Sci. Rep. 6, 22777 (2016).].
We show that decoherence due to the presence of a detector propagates with the speed of light 
 in terms of a retarded Green's function, 
as it should be consistent with causality of relativistic field theories.
The quantum nature of fields, such as quantum fluctuations or emission of 
gravitons expressed in terms of the Keldysh Green's function also play important roles 
in the mechanism of decoherence due to on-shell particle creation. 
We also discuss the trade-off relation between the visibility of the interference and the distinguishability of the measurement, known as
the wave particle duality, in our setup. 
\end{abstract}

\maketitle


\section{Introduction}
Direct detections of gravitational waves from mergers of black holes \cite{LIGOScientific:2016aoc} provide us with solid evidence
that the gravitational interaction is indeed mediated by a gravitational field. 
However, it is not yet experimentally proved that the gravitational field should be quantized. 
Even theoretically, there remains a possibility that gravity is something like an entropic force \cite{Jacobson:1995ab, Verlinde:2010hp}
and not necessarily quantized. So it is becoming more and more important to get any hint 
of the quantum nature of gravity~\cite{Howl:2016ryt}.
The complementarity of quantum mechanics demands that, if gravity is quantized, 
it must show both the particle and wave behaviors such as the photoelectric effect of light or the double-slit experiment of an electron. 
In the double-slit experiment of an electron, the interference pattern of the electron field 
is destroyed if the electron is observed to be localized at one side of the double slit. 
Thus if we can make a coherently superposed state of the gravitational field and then destroy its interference by measurement, 
it becomes a proof of the quantum nature of the gravitational field. 
Bose {\it et al}.-Marletto-Vedral (BMV) experiment was proposed in Refs.~\cite{Bose:2017nin, Marletto:2017kzi}, and many theoretical studies of the 
experiment are given \cite{Miki:2020hvg,Matsumura:2020law, Marshman:2019sne,Christodoulou:2018cmk,Bose:2022uxe,Christodoulou:2022vte,Sugiyama:2022ixw}. 
In particular, the authors in Refs.~\cite{Belenchia:2018szb,Belenchia:2019gcc, Danielson:2021egj} argued how the complementarity of quantum mechanics
is consistent with causality in the relativistic theory by 
discussing how the coherence is casually destroyed by measurement. 

In this paper, we explicitly investigate a slightly modified version of the gedanken experiment 
by Mari and co-workers \cite{Mari:2015qva}
within relativistic quantum field theories as a toy model of the BMV experiment. 
Before explicit calculations, let us recall the basic properties of Green's functions in relativistic quantum field theories. 
Relativistic quantum field theories describe local interactions between sources $J(x)$ of the field $\phi(x)$, 
and various types of Green's functions play different important roles. 
As in  {\it classical} field  theories, the retarded Green's function 
$G_\mathrm{R}(x,y)\coloneqq \ri \theta(t_x-t_y) \expval{[\phi(x), \phi(y)]}$
describes causal influence of a source at $y=(t_y, \vb*{y})$ on the field $\phi(t_x,\vb*{x})$ in future.
Since the field operators $\phi(x)$ and $\phi(y)$ commute if they are separated in the spacelike region, 
the retarded Green's function vanishes there. 
The advanced Green's function  $G_{\mathrm{A}}(x,y)\coloneqq-\ri \theta(t_y-t_x) \expval{ [\phi(x), \phi(y)]}$ is also important as well,
since the theory itself does not distinguish the past and the future. 
The classical electromagnetic (EM) field is usually given by the retarded Li\'{e}nard-Wiechert potential generated by sources of the EM field
as ${\bf A}_\mathrm{R}(x) =\int \dd[4]{y} G_\mathrm{R} (x,y) {\bf J} (y)$.
The advanced potential  is also a solution, and 
 selecting ${\bf A}_\mathrm{R}$ 
requires a specific boundary condition for the homogeneous part of solutions.
Indeed, the retarded potential is selected by imposing a
condition that there are no incoming flux from the past to the volume of our interest (see, e.g., \cite{Davies}, Chap. 5). 

A Green's function specific to {\it quantum} field theories is the
Keldysh Green's function~\cite{Keldysh:1964ud}, $G_\mathrm{K}(x,y)\coloneqq \expval{(\phi(x)\phi(y)+\phi(y)\phi(x))/2}$. 
Its Fourier transform is proportional to  the on-shell $\delta$ function $\delta(k^2+m^2)$, as the field operator $\phi(x)$ creates on-shell states. Thus, this Green's function
appears when we calculate, e.g., an emission rate of radiation. 
The Feynman Green's function is written as a sum $G_\mathrm{F}(x, y) \coloneqq\expval{T\phi(x) \phi(y)}=G_\mathrm{K}-\ri(G_\mathrm{R}+G_{\mathrm{A}})/2$. 
Each Green's function plays a different role. 
In a nutshell, $G_\mathrm{R}$ reflects classical causality and $G_{\mathrm{K}}$ expresses quantum or vacuum fluctuations. 

In this paper, 
we discuss two effects of decoherence on the visibility of interference, one by the Keldish Green's function and the other by the retarded Green's function
in the gedanken experiment as an example.
We show that decoherence by measurement is described by 
the retarded Green's function and propagates with the speed of light, while decoherence by emission 
of on-shell particles is described by the Keldysh Green's function. 
We also discuss the trade-off relation between the visibility and the distinguishability of the measurement, 
known as wave particle duality, in our setup. 
We use the natural units  $\hbar=c=1$.
\\

The paper is organized as follows. In Sec.~\ref{Gedanken}, we explain our setup and summarize the results through three questions and answers. 
In Sec.~\ref{Sec:method}, we introduce our method to calculate various quantities.
Reduced density operators of the system are introduced in Sec.~\ref{Sec:density operator} and then we explain our method of closed time path integral formalism in Sec.~\ref{Sec:CTP}. 
Our results are given in Sec.~\ref{Sec:results}, and 
Sec.~\ref{Sec:conclusion} is devoted to conclusions.
In Appendix~\ref{sec:particle_production}, we prove that the dissipation factor $\Gamma_\mathrm{A}$ is proportional to the created particle number by Alice. In Appendix~\ref{subsec-fieldeffect}, we show the inequality of distinguishabilities $D_\mathrm{B} \le D_{\mathrm{B},\phi}$. 
\section{Gedanken experiment}\label{Gedanken}
\begin{figure}[hbtp]
 \centering
 \includegraphics[keepaspectratio, width=0.6\linewidth]{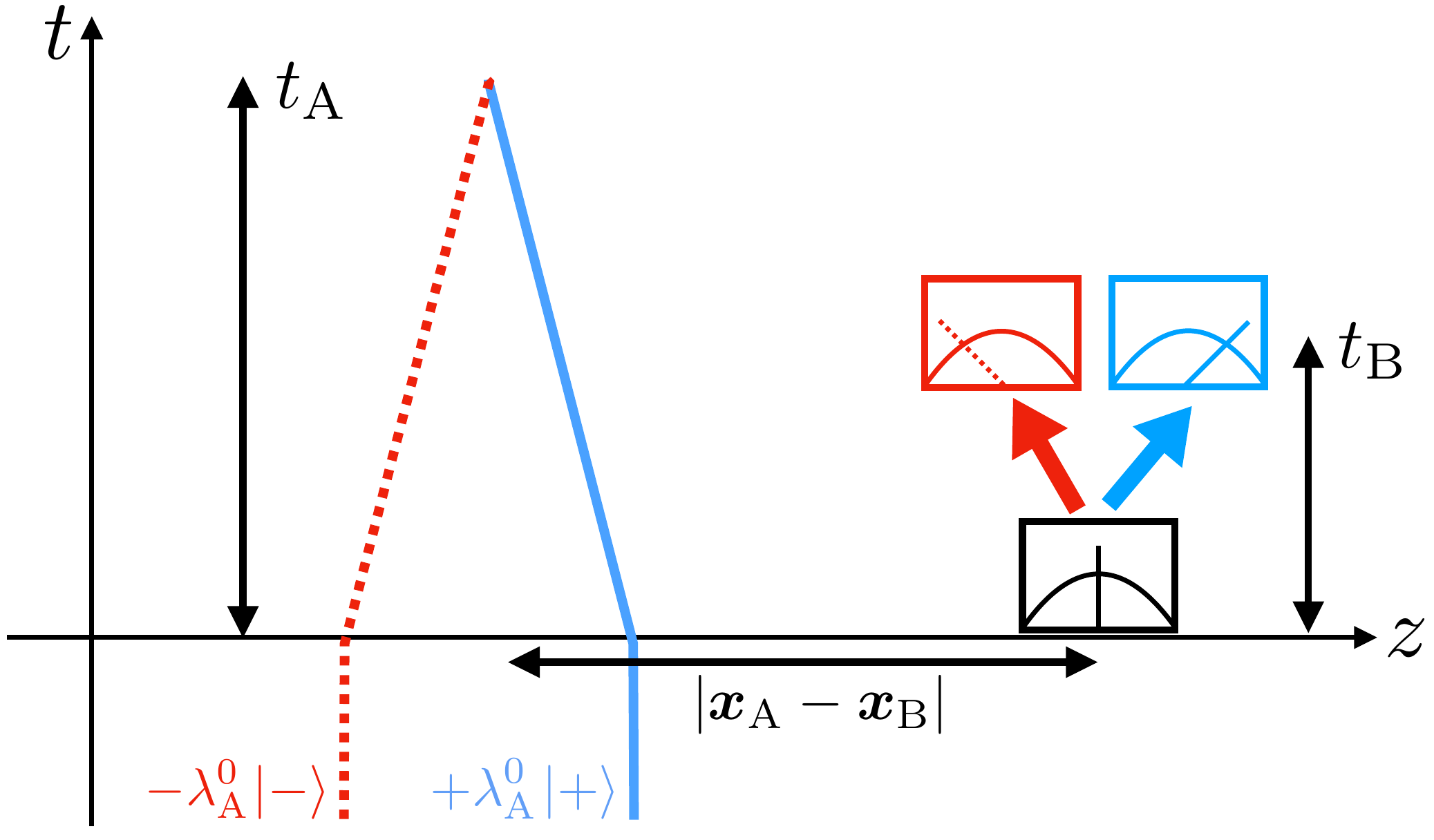}
 \caption{Setup of the gedanken experiment.
 The positions of Alice and Bob are fixed at $\vb*{x}_\mathrm{A}$ and $\vb*{x}_\mathrm{B}$.
 The blue and red-dashed lines schematically represent trajectories of $\lambda_\mathrm{A}(t)\sigma_z$ for $\sigma_z=+1$ and $-1$, respectively.
Bob is equipped with a meter that can measure a value of the field.}
 \label{fig:setup}
\end{figure}
The setup of the gedanken experiment  \cite{Belenchia:2018szb,Belenchia:2019gcc, Danielson:2021egj} shown in Fig.~\ref{fig:setup} is as follows. 
A gravitational field is replaced by a massive scalar field $\phi(x)$ with mass $m$.  
We emphasize that this replacement does not lose important characteristics of the gravitational fields such as entanglement generation mediated by the fields or emission of radiation caused by nonadiabatic processes.
We then 
introduce Alice and Bob at fixed positions $\vb*{x}_\mathrm{A}$ and $\vb*{x}_\mathrm{B}$, respectively.  
Suppose Alice has  a spin  $\sigma_z= \pm 1$, which is 
coupled to the quantum field 
$\phi(x)$ as 
\begin{equation}
H_{\mathrm{A}}= -\lambda_{\mathrm{A}} (t) \sigma_z \phi(t, \vb*{x}_{\mathrm{A}}).
\end{equation}
The time-dependent coupling (protocol) $\lambda_{\mathrm{A}}(t)$ corresponds 
to the  separation of $\sigma_z=\pm 1$ states in the Stern-Gerlach type experiment
of BMV \cite{Bose:2017nin, Marletto:2017kzi}.
Suppose that the initial spin state is given by 
$(\ket{\up}_{\mathrm{A}}  + \ket{\down}_{\mathrm{A}})/\sqrt{2}$, where $\ket{\updown}_{\mathrm{A}}$ correspond to $\sigma_z=\pm 1$ states.
The interference of $\ket{\updown}_{\mathrm{A}}$ is probed
by measuring $\sigma_x$ or $\sigma_y$.  
We choose the protocol $\lambda_{\mathrm{A}}(t)$ as
\eqn{\lambda_{\mathrm{A}}(t)=\bigl(\theta(-t)(1-t/t_{0}) + \theta(t)\theta(t_{\mathrm{A}}-t) (1-t/t_{\mathrm{A}}) \bigr)\lambda_\mathrm{A}^0,
\label{lambdaA}
} 
where $\theta(t)$ is a step function, $\lambda_\mathrm{A}^0$ is a constant, and we take the initial time $t_0 \to - \infty$  at which the vacuum boundary condition is imposed. Alice generates 
a different field configuration corresponding to either $\sigma_z=1$ or $-1$.
Bob observes the value of the field  at $\vb*{x}_{\mathrm{B}}$ by using his   
quantum mechanical variable $\chi_{\mathrm{B}}$ and its conjugate $\pi_{\mathrm{B}}$ satisfying $[\chi_{\mathrm{B}}, \pi_{\mathrm{B}}]=\ri$. 
The coupling of Bob to the field $\phi(x)$ is given by
\begin{equation}
    H_{\mathrm{B}}= \lambda_{\mathrm{B}} (t)\pi_{\mathrm{B}} \phi(t, \vb*{x}_{\mathrm{B}}) .
\end{equation}
The coupling is assumed to be nonzero  $\lambda_{\mathrm{B}}(t)=\alpha$ 
during $t=0$ and $t=t_{\mathrm{B}}>0$.       
Bob's variable $\chi_{\mathrm{B}}$ is shifted by
\begin{equation}
     \chi^{(\phi)}_{\mathrm{B}}(t)=\int^{t} \dd{t^\prime} \lambda_{\mathrm{B}}(t^\prime) \phi(t^\prime, \vb*{x}_{\mathrm{B}}). 
\end{equation}
In this sense, $\chi_{\mathrm{B}}(t)$ is a meter variable that measures  the field configuration at $\vb*{x}_{\mathrm{B}}$.   
Notice that the shifted value $\chi_{\mathrm{B}}^{(\phi)}$ is not a classical number but an operator given in terms of the quantum field $\phi(x).$

Suppose that the $\phi$ field is initially in the ground state.
Because of the coupling  $H_{\mathrm{A}}$, 
a different field configuration is generated corresponding to $\sigma_z=\pm 1$.
Thus the total state of  Alice, the field  $\phi$, and Bob is given by an entangled state,
\al 
{ \ket{\Psi}=
\frac{1}{\sqrt{2}}\qty(\ket{\up}_{\mathrm{A}}  \ket{\Psi_{\up} }_{\phi,\mathrm{B}} + \ket{\down}_{\mathrm{A}} 
\ket{\Psi_{\down}}_{\phi,\mathrm{B}} ) .
\label{total-state}
}
At $t=0$, Bob is not yet coupled to the field and we can write
\eqn
{
\ket{\Psi_{\updown}}_{\phi,\mathrm{B}} =\ket{\Omega_{\updown}}_\phi  
\ket{{B}_0}_{\mathrm{B}} }
where $\ket{{B}_0}_{\mathrm{B}}$ is the initial state of the meter of Bob,
whose wave function $f(\chi)=\bra{\chi}\ket{{B}_0}_{\mathrm{B}}$ is  assumed to have the form
\begin{equation}
f(\chi)\coloneqq(\pi\epsilon^2)^{-\frac{1}{4}}e^{-\frac{\chi^2}{2\epsilon^2}}.
\end{equation}
Here, $\epsilon^2$ represents the variance.
For $t>0$, the coupling $\lambda_{\mathrm{B}} (t)$ is turned on and the meter variable $\chi_{\mathrm{B}}(t)$ is involved in an entangled state with
$
\ket{\Psi_{\updown}}_{\phi,\mathrm{B}} =\ket{\Omega_{\updown} , {B}_{\updown} }_{\phi, \mathrm{B}}.
$
Note that it is not a  tensor product of 
$\ket{\Omega_{\updown}}_\phi$ and $\ket{{B}_{\updown} }_{\mathrm{B}}$.\footnote{If Alice at $t_\mathrm{A}$ and Bob at $t=0$ are spatially separated, both spin states can be transformed to a tensor product state by the same unitary transformation~\cite{Danielson:2021egj} since the time slice of $t_\mathrm{A}$ is deformed to cross $t<0$ point at Bob without using the Hamiltonian $H_\mathrm{A}$, which is 
responsible for vanishing of $\delta_\epsilon(\GRAB)$.}

Under the above setup, we are interested in calculating the interference of spin $\sigma_z=\pm 1$ states called the visibility,
which is given by $v\coloneqq|\braket{ \Psi_\down}{\Psi_\up }_{\phi, \mathrm{B}}| $~\cite{PhysRevA.51.54,PhysRevLett.77.2154,Sugiyama:2022wcd}. 
When the meter of Bob is off, $\sigma_z=\pm 1$ states must be almost identical $|\braket{ \Psi_\down}{\Psi_\up }_{\phi, \mathrm{B}}| \lesssim 1$
as far as the protocol of Alice is sufficiently adiabatic ($t_{\mathrm{A}} \gg 1/m$).
Because of complementarity, if Bob measures to distinguish  the states $\ket{\Psi_{\updown}}_{\phi,\mathrm{B}}$, 
one might expect that the interference disappears
 $|\braket{ \Psi_\down}{\Psi_\up }_{\phi, \mathrm{B}}| \sim 0$ since Bob observes different meter values for $\sigma_z=\pm 1$. 
However, it apparently contradicts the causality if the space-time point of observing $\braket{ \Psi_\down}{\Psi_\up }_{\phi, \mathrm{B}}$ at
$(t_{\mathrm{A}}, \vb*{x}_{\mathrm{A}})$ and that of the measurement at $(0,\vb*{x}_{\mathrm{B}})$ are separated in the spacelike
region. 
The resolution of the paradox is investigated in Refs.~\cite{Belenchia:2018szb,Belenchia:2019gcc, Danielson:2021egj}. 
Focusing on the quantum fluctuations of fields, the authors show that the nonadiabaticity of Alice's protocol
emits on-shell radiation to destroy coherence and this is correlated with an uncertainty of the meter variable
of Bob when $t_\mathrm{A}< |\vb*{x}_{\mathrm{A}}-\vb*{x}_{\mathrm{B}}|$. If the protocol of Alice is adiabatic $v \sim 1$, Bob cannot gain sufficient which-path information. 
In the following, we will investigate the model by an explicit field-theoretic calculation of $\braket{ \Psi_\down}{\Psi_\up }_{\phi, \mathrm{B}}$
and answer the following questions:
\begin{itemize}
 \setlength{\itemsep}{0cm}
\item[Q1] How much decoherence is generated by nonadiabaticity associated with
the Alice protocol $\lambda_{\mathrm{A}}(t)$?
\item[Q2] What meter value does Bob observe? And how much information can Bob get to distinguish Alice's spin?
\item[Q3] How fast does the decoherence by Bob's measurement propagate to Alice?
\end{itemize}

Different properties of Green's functions, $G_\mathrm{R}$, $G_{\mathrm{A}}$, and $G_{\mathrm{K}}$, can answer these questions in a consistent way.  
We use the abbreviations $G^{(pq)}(t,t^\prime)=G((t,\vb*{x}_p),(t^\prime,\vb*{x}_q))$
where $p,q$ are either $\mathrm{A}$, $\mathrm{B}$, or $\vb*{x}$.  
Let us summarize our answers to the above questions. If the field is strongly self-interacting, decoherence by Alice and Bob becomes mixed 
and more complicated. In a weak coupling limit, our calculation shows the following:
\begin{itemize}
 \setlength{\itemsep}{0cm}
\item[A1] 
If  the measurement of  Bob is spatially separated, decoherence is given by $\lambda_{\mathrm{A}}(t)$ as
$\braket{ \Psi_\down}{\Psi_\up }_{\phi, \mathrm{B}}=e^{-\Gamma_{\mathrm{A}}}$, where
\begin{equation}
  \Gamma_{\mathrm{A}} \coloneqq 2\int \dd{t} \int \dd{t^\prime} \lambda_{\mathrm{A}}(t) G_{\mathrm{K}}^{(\mathrm{AA})}(t, t^\prime) \lambda_{\mathrm{A}}(t^\prime).
  \label{eq:GammaA}
\end{equation}
It is due to the emission of on-shell radiation from Alice \cite{Kanno:2020usf}. The number of particles generated by the protocol $\lambda_{\mathrm{A}}(t)$ of Alice is given as $\Gamma_{\mathrm{A}}/2$.
\item[A2] Bob measures the meter value as
$\bra{\Psi_{\updown}}\chi_\mathrm{B}\ket{\Psi_{\updown}}=\pm \overline{\chi_\mathrm{B}}$ for given Alice's spin, where
\begin{equation}
\overline{\chi_\mathrm{B}}=\int \dd{t} \int\dd{t'} \lambda_\mathrm{B}(t) G^{(\mathrm{B}\mathrm{A})}_{\mathrm{R}}(t,t') \lambda_\mathrm{A}(t').\label{eq:chiZB}
\end{equation}
As it is written in terms of the retarded Green's function from Alice to Bob,
it is not directly responsible for the decoherence by Bob's measurement shown in A3.
Nevertheless, it is indirectly related. 
The distinguishability of Bob $D_\mathrm{B}$, defined later in Eq.~\eqref{distinguishability of Bob}, has a trade-off relation with the visibility of Alice $v=|\braket{ \Psi_\down}{\Psi_\up }_{\phi, \mathrm{B}}|$ through the wave particle duality relation
$v^2+D_\mathrm{B}^2 \le 1$ \cite{Sugiyama:2022wcd}. Thus in the adiabatic limit of $v \rightarrow 1$, Bob cannot get any which-path information of Alice. On the other hand, if Bob can distinguish the which-path of Alice, the interference of Alice's 
spin is decohered $v \rightarrow 0$ by an inevitable emission of radiation \cite{Danielson:2021egj}. 
We explicitly calculate $v$ and $D_\mathrm{B}$.
\item[A3] If Bob's measurement is not spatially separated from Alice, interference observed by Alice at time $t (>t_{\mathrm{A}})$ is given by 
$\braket{ \Psi_\down}{\Psi_\up }_{\phi, \mathrm{B}}=e^{-\Gamma_{\mathrm{A}}}  \delta_\epsilon(M)$, where 
\begin{equation}
  \delta_\epsilon (\chi)\coloneqq\exp(-\chi^2/4\epsilon^2) \label{eq:delta_epsilon}
\end{equation}
is an overlap of the wave function of meter,
and 
\begin{equation}
  \GRAB\coloneqq-2 \int \dd{t}\int \dd{t}^\prime \lambda_{\mathrm{A}}(t) G_\mathrm{R}^{(\mathrm{AB})}(t, t^\prime) \lambda_{\mathrm{B}}(t^\prime).
  \label{eq:GRAB}
\end{equation}
Interference is decohered by the causal interaction from $\lambda_{\mathrm{B}}(t)$ to $\lambda_{\mathrm{A}}(t)$, and
described by the retarded Green's function from Bob to Alice. 
Thus if $t_{\mathrm{A}}<|\vb*{x}_{\mathrm{A}}-\vb*{x}_{\mathrm{B}}|$, 
we have $\GRAB=0$ and $\delta_\epsilon(\GRAB) = 1$.  
On the other hand, if $t_{\mathrm{A}}>|\vb*{x}_{\mathrm{A}}-\vb*{x}_{\mathrm{B}}|$,
additional decoherence  $\delta_\epsilon(\GRAB)$ is induced by Bob's measurement.
\end{itemize}

\section{Methods} \label{Sec:method}

Interference of  $\sigma_z=\pm 1$ states can be probed by Alice
 by measuring $\expval{\sigma_x} 
 =\Re \braket{\Psi_\down}{\Psi_\up }_{\phi,\mathrm{B}}$
 and $\expval{{\sigma}_y}=-\Im\braket{\Psi_\down}{\Psi_\up }_{\phi,\mathrm{B}}$.
On the other hand, when Bob reads his meter, it gives either 
$\langle \Psi_\up|\mathrm{\chi_B}| \Psi_\up \rangle_{\phi,\mathrm{B}}$
or $\langle \Psi_\down|\mathrm{\chi_B}| \Psi_\down \rangle_{\phi,\mathrm{B}}$. 
In order to calculate these quantities, we can use the closed time path (CTP) formalism for the field $\phi$. 
The spin variables of Alice and the variable of Bob can be treated in a simpler manner, since the Hamiltonians $H_\mathrm{A}$ and $H_\mathrm{B}$ are diagonalized in terms of  $\sigma_z$ and $\Pi_\mathrm{B}$.

\subsection{Density operator of Alice and Bob}\label{Sec:density operator}
First, since $\sigma_z$ commutes with the Hamiltonian, 
the time evolution is block diagonalized in the $\sigma_z=\pm 1$ basis and the state $\ket{\Psi(t)}$ is given by 
\al 
{ \ket{\Psi(t)}=
\frac{1}{\sqrt{2}}\qty(\ket{\up}_\mathrm{A}  \ket{\Psi_{\up}(t)}_{\phi,\mathrm{B}} + \ket{\down}_\mathrm{A} 
\ket{\Psi_{\down}(t)}_{\phi,\mathrm{B}} ) ,
\label{total-state-t}
}
where each of the $\sigma_z=\pm 1$ states evolves by a unitary operator as
\begin{equation}
    \ket{\Psi_{\updown}(t)}_{\phi,\mathrm{B}}=U(t)_{\sigma_z=\pm 1} \ket{\Psi_{\updown}}_{\phi,\mathrm{B}}.
\end{equation}
Here, the unitary operator is given by $U(t)_{\sigma_z=\pm 1}=\bra{\updown}U(t)\ket{\updown}_\mathrm{A}$ with
the time-evolution operator in the interaction picture,
\begin{equation}
  {U}(t)= T \exp\qty(-\ri\int^t \dd{s}(H_\mathrm{A}(s)+H_\mathrm{B}(s))),
\end{equation}
where $T$ is the time ordering operator.

Furthermore, $\pi_\mathrm{B}$ also commutes with the Hamiltonian, $\ket{\Psi_{\updown}(t)}_{\phi,\mathrm{B}}$ can be written in the form
\begin{equation}
\ket{\Psi_{\updown}(t)}_{\phi,\mathrm{B}} = \int \dd{\Pi}\tilde{f}(\Pi) \ket{\Pi}_{\mathrm{B}} \ket{\Omega_{\updown,\Pi}(t)}_\phi,
\end{equation}
where
\begin{equation}
 \ket{\Omega_{\updown,\Pi}(t)}_\phi =  U_{\sigma_z=\updown, \pi_\mathrm{B}=\Pi}(t)\ket{\Omega}_\phi
\end{equation} 
is the state of field $\phi$ for fixed Alice's spin and Bob's momentum, and
$\tilde{f}(\Pi)=\bra{\Pi}\ket{B_0}_\mathrm{B}$ is the wave function of Bob in the momentum space,
\begin{equation}
  \tilde{f}(\Pi)= \int\frac{\dd{\chi}}{\sqrt{2\pi}} e^{-\ri \Pi\chi} f(\chi) =  \qty(\frac{\epsilon^2}{\pi})^{\frac{1}{4}}e^{-\frac{\epsilon^2}{2}\Pi^2} .
\end{equation} 

In many cases, we are interested in the degrees of freedom of Alice and Bob, and it is convenient to trace out the field degrees of freedom.
For this purpose, we introduce the reduced density operator,
\begin{align}
  \rho_{\mathrm{A},\mathrm{B}}&\coloneqq
  \tr_\phi \ket{\Psi(t)}\bra{\Psi(t)}\notag\\
   &= 
  \sum_{\sone,\stwo={\updown}}\int\dd{\Pione}\int\dd{\Pitwo}
  \frac{\tilde{f}(\Pione)\tilde{f}^*(\Pitwo)}{2}
  \bra*{\Omega_{\stwo,\Pitwo}(t)} \ket*{\Omega_{\sone,\Pione}(t)} 
  \ket{\sone}_\mathrm{A}\ket{\Pione}_\mathrm{B}\bra{\Pitwo}_\mathrm{B}\bra{\stwo}_\mathrm{A}.
\end{align}
Similarly, the reduced density operators of Alice or Bob are obtained by taking further traces over Bob or Alice, given by 
\begin{align}
  \rho_{\mathrm{A}}&\coloneqq
  \tr_{\phi,\mathrm{B}} \ket{\Psi(t)}\bra{\Psi(t)}\notag\\
   &= 
  \sum_{\sone,\stwo={\updown}}\int \dd{\Pi}
  \frac{|\tilde{f}(\Pi)|^2}{2}
  \bra*{\Omega_{\stwo,\Pi}(t)}\ket*{\Omega_{\sone,\Pi}(t)}
  \ket{\sone}_\mathrm{A}  \bra{\stwo}_\mathrm{A},
  \label{eq:density matrix:sigma}
\end{align}
and
\begin{align}
  \rho_{\mathrm{B}}&\coloneqq
  \tr_{\mathrm{A},\phi} \ket{\Psi(t)}\bra{\Psi(t)}
   = \frac{1}{2}\rho^{\up}_\mathrm{B}+\frac{1}{2}\rho^{\down}_\mathrm{B}.
\end{align}
In the last line, we decompose $\rho_{\mathrm{B}}$ by introducing
\begin{equation}
  \rho^{\updown}_\mathrm{B}=
  \tr_\phi \ket{\Psi_{\updown}(t)}\bra{\Psi_{\updown}(t)}=
  \int\dd{\Pione}\int\dd{\Pitwo}
  \tilde{f}(\Pione)\tilde{f}^*(\Pitwo)
  \bra*{\Omega_{\updown,\Pitwo}(t)} \ket*{\Omega_{\updown,\Pione}(t)} 
  \ket{\Pione}_\mathrm{B}\bra{\Pitwo}_\mathrm{B}, 
  \label{eq:rho_updown}
\end{equation}
each of which corresponds to $\sigma_z=\pm 1$.
For explicit calculations, we 
need the computation of the inner product of the field, $\bra*{\Omega_{\stwo,\Pitwo}(t)} \ket*{\Omega_{\sone,\Pione}(t)}$. 
This can be done in the next section using the closed time path formalism 
often used in nonequilibrium quantum field theories~\cite{Schwinger:1960qe,Keldysh:1964ud,Rammer:2007zz}.
By using the density operators, we can obtain  
$\expval{\sigma_{x,y}}=\tr\rho_\mathrm{A}\sigma_{x,y}$
or 
$\bra{\Psi_\updown}\chi_\mathrm{B}\ket{\Psi_\updown}_{\phi,\mathrm{B}}=
\tr \rho_\mathrm{B}^{\updown}\chi_\mathrm{B}$ .

\subsection{Closed time path formalism}\label{Sec:CTP}
The inner product $\bra*{\Omega_{\stwo,\Pitwo}(t)} \ket*{\Omega_{\sone,\Pione}(t)} $ can be calculated by using the technique of the CTP formalism.
Since $U_{\sigma_z=\si, \pi_\mathrm{B}=\Pii}$ represents the time evolution of fields at fixed $\sigma_z$ and $\pi_\mathrm{B}$, $\bra{\phi} U_{\sigma=\si, \pi_\mathrm{B}=\Pii}(t)\ket{\Omega}$ can be written in the Feynman path integral form as
\begin{equation}
  \bra{\phi} U_{\sigma_z=s_i, \pi_\mathrm{B}=\Pi_i}(t)\ket{\Omega}
  =\int\mathcal{D}\phi_i e^{\ri S[\phi_i,J_i]},
\end{equation}
where the action is 
\begin{equation}
  S[\phi_i,J_i] = \int\dd[4]{x}\qty(
    \frac{1}{2}(\partial_t\phi_i(x))^2-\frac{1}{2}
    (\vb*{\nabla}\phi_i(x))^2-\frac{1}{2}m^2(\phi_i(x))^2
    +\phi_i(x) J_i(x)
  ).
  \label{eq:action}
\end{equation}
Here, the Hamiltonians of Alice and Bob are treated as the source term, $J_i =J_i^\mathrm{A}+J_i^{\mathrm{B}}$, with
\al{
J_i^\mathrm{A}(x) &=  \lambda_{\mathrm{A}}(t) \si \delta^{(3)}(\vb*{x}-\vb*{x}_{\mathrm{A}}),\label{eq:J_A}
\\
J_i^\mathrm{B}(x) &= -\lambda_{\mathrm{B}}(t)\Pii \delta^{(3)}(\vb*{x}-\vb*{x}_{\mathrm{B}}).
}
Using this path integral formula, we can also express
the inner product $\bra*{\Omega_{\stwo,\Pitwo}(t)} \ket*{\Omega_{\sone,\Pione}(t)} $  as
the path integral with both the forward and backward time paths,
\al{
  \bra*{\Omega_{\stwo,\Pitwo}(t)} \ket*{\Omega_{\sone,\Pione}(t)} 
&=
\int \dd\phi\bra{\Omega}U^\dag_{\sigma=\stwo, \pi_\mathrm{B}=\Pitwo}(t)\ket{\phi}\bra{\phi}
U_{\sigma=\sone, \pi_\mathrm{B}=\Pione}(t)\ket{\Omega}\notag\\
&=\int \mathcal{D}\phi_1  \mathcal{D}\phi_2 
e^{\ri(S[\phi_1,J_1]-S [\phi_2,J_2])}.
\label{U1U2}
}
Here, the integral of $\phi$ at the final state is included in the path integral, and the boundary condition, $\phi_1(t,\vb*{x})=\phi_2(t,\vb*{x})$ is imposed.
On the other hand, the vacuum boundary condition is imposed for the initial state, which corresponds to no incoming flux condition from the past for the retarded Li\'{e}nard-Wiechert potential in the
classical EM field.

Equation~\eqref{U1U2} is a Gaussian integration, 
so that the path integral can be  evaluated by using Green's functions as
$\bra*{\Omega_{\stwo,\Pitwo}(t)} \ket*{\Omega_{\sone,\Pione}(t)}=\exp\qty(\mathrm{i}W[J_1,J_2])$~\cite{Rammer:2007zz}, where
\al{
\ri W[J_1,J_2]= -\frac{1}{2} \int
\dd[4]{x} \dd[4]{y}{J}^{i}(x) G_{ij}(x,y) {J}^{j}(y),
\label{eq:generating_functional}
}
is the generating functional.
Here, currents with upper indices are defined as $(J^1,J^2)\coloneqq(J_1, -J_2)$,
which reflect the negative sign of the backward path in Eq.~\eqref{U1U2}.
Green's functions $G_{ij}$ are given by 
$G_{11}(x,y)=G_\mathrm{F}(x,y)$, $G_{12}(x,y)=\expval{\phi(y) \phi(x)}$, and
$G_{21}(x,y)=\expval{\phi(x) \phi(y)}$,
and $G_{22}(x,y)=G_{\bar{\mathrm{F}}} (x,y) $
is the antitime ordered product.

It is useful to recombine fields and currents as 
\begin{align}
& \phi_{\mathrm{r}}=\frac{1}{2}(\phi_1+\phi_2), \ \phi_{\mathrm{a}}=(\phi_1-\phi_2), \nonumber \\
& J^{\mathrm{a}}=J_{\mathrm{r}}=\frac{1}{2}(J_1+J_2), \ J^{\mathrm{r}}=J_{\mathrm{a}}=(J_1-J_2).
 \end{align}
 In this new basis, the Green's functions are 
 given as follows: First, $G_\mathrm{rr}(x,y)=G_{\mathrm{K}}(x,y)$ is the Keldysh Green's function, which describes quantum 
 fluctuations and emission of on-shell particles. Second, $G_\mathrm{ra}(x,y)=-\ri G_\mathrm{R}(x,y)$ is the retarded Green's function.
 $G_\mathrm{ar}(x,y)=-\ri G_{\mathrm{A}}(x,y)=-\ri G_\mathrm{R}(y,x)$ is similar. They describe causal processes
 since they vanish in the spacelike region. Finally, $G_\mathrm{aa}(x,y)=0$. It follows from the unitarity 
 that Eq.~\eqref{U1U2} becomes one if $\sone=\stwo$ and $\Pione=\Pitwo$.
 
To summarize, $W$ is written  as
\al{
\ri W=\int\dd[4]{x}\dd[4]{y}
\left[-\frac{1}{2}{{J}_{\mathrm{a}}(x)G_{\mathrm{K}}(x,y) {J}_{\mathrm{a}} (y)}+\ri {J}_{\mathrm{a}}(x)G_\mathrm{R}(x,y) {J}_{\mathrm{r}}(y)\right],
\label{finalW}
}
and the sources  by Alice and Bob are given by
\begin{align}
J_{\mathrm{r}}^\mathrm{A} &= \lambda_{\mathrm{A}}(t)  \sr \delta^{(3)}(\vb*{x}-\vb*{x}_{\mathrm{A}}), \  \
J_{\mathrm{a}}^\mathrm{A} =  \lambda_{\mathrm{A}}(t)\sa \  \delta^{(3)}(\vb*{x}-\vb*{x}_{\mathrm{A}}) , \nonumber  \\
J_{\mathrm{r}}^\mathrm{B} &= -\lambda_{\mathrm{B}}(t)\Pir  \delta^{(3)}(\vb*{x}-\vb*{x}_{\mathrm{B}}) , \  \  J_{\mathrm{a}}^\mathrm{B}=
-\lambda_{\mathrm{B}}(t) \Pia \delta^{(3)}(\vb*{x}-\vb*{x}_{\mathrm{B}}),
\end{align}
where $\sr=(\sone + \stwo)/2 $, $\sa= \sone - \stwo $, $\Pir=(\Pione+\Pitwo)/2$, and $\Pia=\Pione-\Pitwo$.
In Ref.~\cite{Christodoulou:2022vte}, the authors discuss causal entanglement generation based on the second term in Eq.~\eqref{finalW},
which they call on-shell action. 
As far as the propagation of entanglement generation or decoherence is concerned, it gives a correct answer. For discussing quantum emission 
of particles like $\Gamma_{\mathrm{A}}$, the first term is necessary in addition to the on-shell action.
For later use, it may be convenient to express $\ri W$ as the following explicit form,
\begin{equation}
  \begin{split}
    \ri W
  &= -\frac{\sa^2}{4}\Gamma_\mathrm{A} -\frac{\Pia^2}{4} \Gamma_\mathrm{B}
  -  \frac{\Pia\sa}{2} \Gamma_{\mathrm{AB}}
  +\ri \Pia \Pi_\mathrm{r} \mathfrak{G}_\mathrm{R}^{\mathrm{BB}} 
   -\ri\Pia   \sr \overline{\chi_\mathrm{B}}
   +\ri   \frac{\Pir\sa}{2} \GRAB,
  \end{split}
\end{equation}
where we define
\begin{align}
  \Gamma_\mathrm{B}&=2\int \dd{t} \int\dd{t'} \lambda_\mathrm{B}(t) G^{(\mathrm{B}\mathrm{B})}_{\mathrm{K}}(t,t')\lambda_\mathrm{B}(t'),\\
  \Gamma_{\mathrm{AB}}&=-2\int \dd{t} \int\dd{t'} \lambda_\mathrm{A}(t) G^{(\mathrm{A}\mathrm{B})}_{\mathrm{K}}(t,t')\lambda_\mathrm{B}(t')  ,\\
  \mathfrak{G}_\mathrm{R}^{\mathrm{BB}}&=\int \dd{t} \int\dd{t'}\lambda_\mathrm{B}(t) G^{(\mathrm{B}\mathrm{B})}_{\mathrm{R}}(t,t') \lambda_\mathrm{B}(t').
\end{align}
$\Gamma_\mathrm{A}$, $\overline{\chi_\mathrm{B}}$ and $\GRAB$ are defined in Eqs.~\eqref{eq:GammaA}, \eqref{eq:chiZB}, and \eqref{eq:GRAB}, respectively.
As will be shown, $\overline{\chi_\mathrm{B}}$ is the meter's expectation value when the spin of Alice is given by $\sigma_z=+1$.
For $\sigma_z=-1$, the meter value is given by its opposite $- \overline{\chi_\mathrm{B}}$.

\section{Results}\label{Sec:results}
\subsection{Observables of Alice and Bob}
 Now we are ready to calculate various quantities such as  
$\braket{\Psi_\down}{\Psi_\up}_{\phi,\mathrm{B}}=\tr_\mathrm{A} \rho_\mathrm{A}(\sigma_x-\ri\sigma_y)=\bra{+}\rho_\mathrm{A}\ket{-}$ or  a meter value $\bra{\Psi_{\updown}}\chi_\mathrm{B}\ket{\Psi_{\updown} }_{\phi,\mathrm{B}}=\tr_\mathrm{B}\rho^{\updown}_B\chi_\mathrm{B}$.
For calculating the interference $\braket{ \Psi_\down}{\Psi_\up }_{\phi, \mathrm{B}}$, we set $\sone=+$, $\stwo=-$, and $\Pione=\Pitwo=\Pir$.
 By integrating $\Pir$, we have
\al{
\braket{ \Psi_\down}{\Psi_\up }_{\phi, \mathrm{B}} = \int\dd\Pir
|\tilde{f}(\Pir)|^2
e^{\ri W}
= \sqrt{\frac{\epsilon^2}{\pi}}\int\dd\Pir
e^{-\epsilon^2\Pir^2-\Gamma_{\mathrm{A}}+\ri\Pir \GRAB}
 =e^{-\Gamma_{\mathrm{A}}} \delta_\epsilon (\GRAB),
\label{result-c}
}
where $\Gamma_{\mathrm{A}}$, $\GRAB$, and $\delta_\epsilon(M)$ are defined before in Eqs.~\eqref{eq:GammaA} and \eqref{eq:delta_epsilon}. 
$\Gamma_{\mathrm{A}}$ contains the Keldysh Green's function connecting the protocol $\lambda_{\mathrm{A}}(t)$ of Alice. 
It can be estimated as
\al{
\Gamma_{\mathrm{A}}=\int \frac{\dd[3]{k}}{(2\pi)^3}\frac{2(\lambda_\mathrm{A}^0)^2}{t_{\mathrm{A}}^2\omega_{\vb*{k}}^5} \qty(1-\cos(t_{\mathrm{A}} \omega_{\vb*{k}})),
}
where $\lambda_\mathrm{A}^0$ is the magnitude of $\lambda_{\mathrm{A}}$ defined in Eq.~\eqref{lambdaA}. 
Since it contains the Keldysh Green function, it represents decoherence caused by the emission of 
on-shell radiation from Alice together with quantum fluctuation of fields in the vacuum. 
The quantity $\Gamma_{\mathrm{A}}/2$ is nothing but the number of created particles in the weak coupling limit, whose derivation is shown in Appendix.~\ref{sec:particle_production}.
In the adiabatic limit of $t_{\mathrm{A}} \rightarrow \infty$, $\Gamma_{\mathrm{A}} \rightarrow 0$ and the decoherence by Alice disappears, as far as 
$M=0$ in the spacelike separated case discussed below. 

 Alice observes $\sigma_{x,y}$ to 
probe the interference  at $(t, \vb*{x}_{\mathrm{A}})$, and Bob sets his meter on 
at $(t=0, \vb*{x}_{\mathrm{B}})$. Since $\GRAB$ contains the retarded Green's function from Bob to Alice, 
the decoherence due to $\GRAB$ is generated by causal interaction\footnote{Causality is reflected in commutators of separated operators in the Hamiltonian formalism. See, e.g.,  Ref.~\cite{PhysRevD.92.104019}}
from $\lambda_{\mathrm{B}}(t)$ to $\lambda_{\mathrm{A}}(t)$.
Thus, if  $t_{\mathrm{A}}< |\vb*{x}_{\mathrm{A}}-\vb*{x}_{\mathrm{B}}|$, 
the retarded Green's function vanishes and $M=0$. 
As seen below, 
 the interference of Alice's spin is not affected by the measurement by Bob due to causality.
The property $\delta_\epsilon (\GRAB)=1$  does hold
 even when Alice observes $\sigma_{x,y}$  at a sufficiently far future
when it is causally connected to $\lambda_{\mathrm{B}}(t)$. 
It is because Alice's spin system is already decoupled from the field $\phi(x)$ after $t=t_{\mathrm{A}}$ when it is not yet 
causally connected to Bob's measurement. 
On the other hand, if the information of the measurement by Bob arrives  earlier than $t_{\mathrm{A}}$,
$\delta_\epsilon (\GRAB)$ can be smaller than 1 so that Bob decoheres the interference of Alice's spin. 
$\GRAB$ can be large if Bob is near Alice and sets his meter on much earlier than $t_{\mathrm{A}}$.
Simultaneously we can take  an adiabatic limit of large $t_{\mathrm{A}}$ so that $e^{-\Gamma_{\mathrm{A}}} \sim 1$.
Thus, decoherence $e^{-\Gamma_{\mathrm{A}}}$ caused by the nonadiabaticity of Alice
 and $\delta_\epsilon (\GRAB)$ by the presence of the meter at Bob are independent as far as the self-interaction of fields is neglected.
 
 Let us consider Bob's measurement when Alice observes $\sigma_z=\pm1$. 
 Suppose that Alice and Bob are spatially separated. The conditional probability that Bob observes his meter value as 
 $\chi_\mathrm{B}=\chi$  for a given Alice's observable $\sigma_z=\updown$ is defined by
 \begin{equation}
  P^\updown(\chi)\coloneqq 
  \frac{\tr_{\mathrm{A},\mathrm{B}} \rho_{\mathrm{A},\mathrm{B}}\ket{\updown}_\mathrm{A}\ket{\chi}_\mathrm{B}\bra{\chi}_\mathrm{B}\bra{\updown}_\mathrm{A}}{
 \tr_{\mathrm{A},\mathrm{B}}\rho_{\mathrm{A},\mathrm{B}}\ket{\updown}_\mathrm{A}\bra{\updown}_\mathrm{A}
  }=\tr_\mathrm{B} \rho^{\pm}_{\mathrm{B}}\qty(\ket{\chi}_\mathrm{B}\bra{\chi}_\mathrm{B}),
 \end{equation}
which is  calculated as
\begin{align}
  P^{\updown}(\chi)
  &= \int\dd{\Pione}\dd{\Pitwo}
  \tilde{f}(\Pione)\tilde{f}^*(\Pitwo)
  \frac{e^{\ri\chi(\Pione-\Pitwo)}}{2\pi}
  e^{\ri W}\notag\\
  &=\frac{1}{2\pi}
\sqrt{\frac{\epsilon^2}{\pi}}
  \int\dd{\Pi_\mathrm{r}}\dd{\Pi_\mathrm{a}}
  e^{-\epsilon^2\Pi_\mathrm{r}^2-\frac{\epsilon^2+\Gamma_\mathrm{B}}{4}\Pi_\mathrm{a}^2+\ri(\chi\mp\overline{\chi_\mathrm{B}}) \Pi_\mathrm{a}}
  e^{ 
  +\ri \Pia \Pi_\mathrm{r} \mathfrak{G}_\mathrm{R}^{\mathrm{BB}} 
   }.
\end{align}
Again, this is a Gaussian integral, and we can easily perform the integration to obtain
\begin{equation}
  P^\updown(\chi)=
  \frac{1}{\sqrt{2\pi  \Sigma^2}}
    \exp\qty(-\frac{1}{2\Sigma^2}\qty(\chi\mp\overline{\chi_\mathrm{B}} )^2),
    \label{eq:P(s,chi)}
\end{equation}
where
\begin{equation}
  \Sigma^2 =\frac{\qty(\mathfrak{G}_\mathrm{R}^{\mathrm{BB}})^2}{2\epsilon^2}+\frac{1}{2}(\Gamma_\mathrm{B} +\epsilon^2).
  \label{eq:Sigma}
\end{equation}
The conditional probability is peaked around the value $\chi = \pm \overline{\chi_\mathrm{B}}$ with variance $\Sigma^2$.

Let us explicitly calculate the meter value
$\tr_\mathrm{B} \rho^{\updown}_\mathrm{B}\chi_\mathrm{B}$.
It is given by
\al{
\tr_\mathrm{B} \rho^{\updown}_\mathrm{B}\chi_\mathrm{B}&=\int\dd{\chi} \chi P^\updown(\chi)
= \updown\overline{\chi_\mathrm{B}}
=\pm\int \dd{t} \int\dd{t'} \lambda_\mathrm{B}(t) G^{(\mathrm{B}\mathrm{A})}_{\mathrm{R}}(t,t') \lambda_\mathrm{A}(t').
\label{field-vev}
}
Thus the expectation value is determined by the source at Alice in the causal past of Bob. 
Similarly, the variance is given by 
\begin{align}
  \tr_\mathrm{B} \rho^{\updown}_\mathrm{B}(\chi_\mathrm{B}\mp\overline{\chi_\mathrm{B}})^2
  =\int\dd{\chi} (\chi\mp\overline{\chi_\mathrm{B}})^2 P^\updown(\chi)
  =\Sigma^2.
\label{eq:variance}
\end{align}
The variance represents the uncertainty of the measurement.
From Eq.~\eqref{eq:Sigma}, we can see that 
there are three contributions to the variance,
$\epsilon^2/2$, $\Gamma_\mathrm{B}/2$, and ${(\mathfrak{G}_\mathrm{R}^{\mathrm{BB}})^2/}{2\epsilon^2}$.
$\epsilon^2/2$ represents the variance of the meter's wave function that determines the sensitivity of the meter.
$\Gamma_\mathrm{B}/2$ represents the number of created particles by Bob's protocol $\lambda_\mathrm{B}(t)$, which causes the additional uncertainty of the measurement. $\mathfrak{G}_\mathrm{R}^{\mathrm{BB}}$ represents the self-correlation through the field.
Obviously, there is the optimal value of $\Sigma^2$, which is obtained from $\mathrm{d}\Sigma^2/\mathrm{d}\epsilon^2=0$,
\begin{equation}
\epsilon^2= |\mathfrak{G}_\mathrm{R}^{\mathrm{BB}}|,
\end{equation}
and the optimal value is
\begin{equation}
  \Sigma^2 =|\mathfrak{G}_\mathrm{R}^{\mathrm{BB}} |+\frac{1}{2}\Gamma_\mathrm{B}.
\end{equation}
Within our measurement model, this result implies that there are unavoidable uncertainties of measurement due to the interaction with the field.
A large uncertainty prevents Bob from distinguishing Alice's spin.
In the next subsection, we discuss the distinguishability of Bob $D_\mathrm{B}$ and the relation between it and the visibility of Alice $v$,
which is given by the inequality 
$v^2+D_\mathrm{B}^2 \leq 1$.

\subsection{Visibility of Alice and distinguishability of Bob}\label{Sec:visibility}

An inequality can be proved among the various quantities of Alice and Bob, such as the visibility (related to expectation values of 
Alice's spin $\sigma_x$ and $\sigma_y$) and the distinguishability (related to the magnitude of Bob's meter value).
For this, we introduce 
the following reduced density operator: 
\begin{equation}
   \rho_{\mathrm{B},\mathrm{\phi}}^{\updown} \coloneqq  \ket{\Psi_\updown(t)}_{\phi,\mathrm{B}} \bra{\Psi_\updown(t)}_{\phi,\mathrm{B}}  ,
\end{equation}
describing the state of Bob and field,
in addition to Eq.~\eqref{eq:rho_updown}.
The off-diagonal element $ \braket{\Psi_\down(t)}{\Psi_\up(t)}_{\phi,\mathrm{B}}$
is measured by observing $\sigma_{x,y}$ and related to 
the interference of spin-up and spin-down wave functions of Alice. 
On the other hand, each diagonal element of spin-up  $ \bra{\Psi_\up(t)}\chi_\mathrm{B}\ket{\Psi_\up(t)}_{\phi,\mathrm{B}}$
or spin-down $ \bra{\Psi_\down(t)}\chi_\mathrm{B}\ket{\Psi_\down(t)}_{\phi,\mathrm{B}}$ 
gives the meter value of Bob when Alice measures spin-up or spin-down. 

An inequality known as the wave particle duality gives a trade-off relation of Alice's visibility
and distinguishability of Bob's measurement. 
We first review the proof of the inequality from a general argument. 
Our system is composed of the states of Alice, Bob, and the field $\phi$, and
the state of the total system is given as in Eq.~\eqref{total-state}. 
Visibility of Alice is defined by~\cite{PhysRevA.51.54,PhysRevLett.77.2154,Sugiyama:2022wcd}
\begin{eqnarray}
v^2 &\coloneqq& \langle \sigma_x \rangle ^2 + \langle \sigma_y \rangle ^2
= | \braket{\Psi_\down(t)}{\Psi_\up(t)}_{\phi,\mathrm{B}}|^2  .
\end{eqnarray}
On the other hand, the distinguishability of Bob is defined by the trace distance between 
$\rho_\mathrm{B}^{\up}$ and $\rho_\mathrm{B}^{\down}$,
\begin{equation}
    D_\mathrm{B} \coloneqq \frac{1}{2} \tr_\mathrm{B} |\rho_\mathrm{B}^{\up} - \rho_\mathrm{B}^{\down}|.
    \label{distinguishability of Bob}
\end{equation}
Bob can distinguish Alice's spin from the difference of density operators $\rho_\mathrm{B}^{\up}$ and $\rho_\mathrm{B}^{\down}$.
Some of the information of Alice's spin is lost in configurations of the fields. 
Suppose that we can perform a complete measurement by using the reduced density operator $\rho_{\mathrm{B},\mathrm{\phi}}^{\updown}$. 
Then defining the trace distance of $\rho_{\mathrm{B},\mathrm{\phi}}^{\updown}$,
\begin{equation}
  D_\mathrm{B,\phi} \coloneqq \frac{1}{2}   \tr_{\mathrm{B},\phi} |\rho_{\mathrm{B},\phi}^{\up} - \rho_{\mathrm{B},\phi}^{\down}|,
\end{equation}
the equality holds
\begin{equation}
    v^2+D_{\mathrm{B},\phi}^2=1 ,
    \label{WDequality}
\end{equation}
since the density operator $\rho_{\mathrm{B},\phi}^{\pm}$ describes a pure state.  
Indeed, 
we can restrict the Hilbert space of Bob and the field into a two-dimensional subspace
spanned by $\ket{\Psi_\up(t)}_{\phi,\mathrm{B}}$
and $\ket{\Psi_\down(t)}_{\phi,\mathrm{B}}$. Then the nonzero eigenvalues of the operator
$(\rho_{\mathrm{B},\phi}^{\up} - \rho_{\mathrm{B},\phi}^{\down})$ are given by 
$\pm \sqrt{1-|\braket{\Psi_\down(t)}{\Psi_\up(t)}_{\phi,\mathrm{B}}|^2}$, and Eq.~\eqref{WDequality} does hold.
A proof is given in Appendix~\ref{subsec-fieldeffect}. 

If we abandon the information of the field and use the limited information of Bob, described by distinguishability $D_\mathrm{B}$ instead of $D_{\mathrm{B}, \phi}$, 
we have  
\begin{equation}
    D_\mathrm{B} 
    \leq D_{\mathrm{B},\phi}, 
\end{equation}
as shown in Appendix.~\ref{subsec-fieldeffect}.
Thus, we have an inequality
\begin{equation}
    v^2+D_\mathrm{B}^2 \leq 1,
    \label{WD-Bob}
\end{equation}
which gives a trade-off relation between the visibility of Alice and the distinguishability of Bob's measurement. 
The equality holds when the effect of the field does not further mix the density operators $\rho_\mathrm{B}^{\pm}$. 
This inequality is used to show the complementarity of Alice's interference and Bob's measurement
by Ref.~\cite{Sugiyama:2022wcd} in a case where Bob is also described by a spin 1/2 system.
In the following, we apply this inequality to our system where Bob is described by a continuous variable $\chi_\mathrm{B}$.
In order to evaluate $D_\mathrm{B}$, we use the following property of the trace distance~\cite{10.5555/1972505}:
\begin{equation}
    D_\mathrm{B} = \text{max}_{P} \left[ \left|\tr_\mathrm{B} P (\rho_\mathrm{B}^{\up} - \rho_\mathrm{B}^{\down})\right| \right]
    \coloneqq  \text{max}_{P} D_\mathrm{B}^{(P)} = D_\mathrm{B}^{(P_>)},
    \label{WPduality-ours}
\end{equation}
where a maximum value is taken over all possible projection operators $P$.
The projector maximizing $D_\mathrm{B}^{(P)}$ is given by the projection operator $P_>$ on the vector space spanned by all the  eigenstates
with a positive  eigenvalue of $(\rho_\mathrm{B}^{\up} - \rho_\mathrm{B}^{\down})$ or, equivalently, on 
the states with negative eigenvalues $P_<$.
For a general projector, 
the inequality $v^2+ (D_\mathrm{B}^{(P)})^2 \leq 1$ is satisfied.
The visibility of Alice, the first term of Eq.~\eqref{WPduality-ours}, can take one in the adiabatic limit. 
In the limit, Bob cannot distinguish Alice's spin since the second term must vanish.
On the other hand, if Bob can distinguish Alice's spin, the difference of field values at Bob generated by
Alice's spin-up or spin-down becomes sufficiently large. Then 
the second term approaches one and the visibility vanishes. 
Physically, the vanishing of visibility is associated with the emission of on-shell particles. 
Namely, in order to transmit the information from Alice to Bob, Alice must emit a sufficient amount of 
on-shell particles into the open space.

\subsection{Complementarity}\label{Sec:complementariry}
Let us now apply the wave particle duality to our case. 
The visibility is given by 
\begin{equation}
   v^2 =
 \qty[ e^{-\Gamma_\mathrm{A}}  
 \delta_\epsilon(\GRAB) ]^2.
\end{equation}
The distinguishability is obtained by calculating the eigenvalues of the operator
\begin{eqnarray}
\tr_\mathrm{A}\rho_{\mathrm{A},\mathrm{B}}\sigma_z= \frac{1}{2}\rho_\mathrm{B}^{\up} -  \frac{1}{2}\rho_\mathrm{B}^{\down}.
\end{eqnarray}
The trace distance between these two density operators can be obtained by diagonalizing the operator $(\rho_\mathrm{B}^{\up}-\rho_\mathrm{B}^{\down})$
in the Hilbert space of Bob. Here we evaluate the trace distance by choosing an appropriate projector $P_h$ as follows.
The projector we choose is 
\begin{eqnarray}
 P_h \coloneqq \int \dd\chi h(\chi) \ket{\chi}\bra{\chi},
\end{eqnarray}
where $\ket{\chi}$ is an eigenstate of ${\chi}_\mathrm{B}$ representing the meter value of Bob, and $h(\chi)$ is a variational function with the property of a projector $h(\chi)=0$ or $1$.
Namely, the function $h(\chi)$ is a step function taking value $1$ in  specific regions, and $0$ otherwise. 
Then the trace distance $ D_{\mathrm{B}}^{(P_h)}$ evaluated by this variational projector is given by
\begin{eqnarray}
 D_{\mathrm{B}}^{(P_h)} = \left| \int d\chi h(\chi) \left( P^+(\chi) -P^-(\chi) \right) \right|.
\end{eqnarray}
Since $P^{\updown}(\chi)$ are Gaussian given in Eq.~\eqref{eq:P(s,chi)} with the center $\chi=\mp\overline{\chi_\mathrm{B}} $, the maximum value of $D_\mathrm{B}^{(P_h)}$
is given by $h(y)=\theta(y)$. Thus, we have the maximum value of $D_{\mathrm{B}}^{(P_h)}$ among the above projectors 
\begin{equation}
  D_\mathrm{B}^{(P_\theta)} = \left| \int_0^\infty  \dd{\chi} \left( P^+(\chi) -P^-(\chi) \right) \right| 
  =\left| \mathbf{erf}\qty(\frac{\overline{\chi_\mathrm{B}}}{\sqrt{2}\Sigma} ) \right|,
\end{equation}
where the error function is given by
\begin{eqnarray}
  \mathbf{erf}(x) \coloneqq 2 \int_0^{x} \frac{dt}{\sqrt{\pi}} e^{-t^2}.
\end{eqnarray}
The error function is monotonically increasing, and is bounded in $-1 \le \mathbf{erf}(x) \le 1$.

From the wave particle duality of Alice's and Bob's measurements, we have the inequality
\begin{eqnarray}
 \Bigl[e^{-\Gamma_\mathrm{A}}  
 \delta_\epsilon{\qty(\GRAB)} \Bigr]^2  + \Bigl[ \mathbf{erf}\Bigl(\frac{\overline{\chi_\mathrm{B}}}{\sqrt{2}\Sigma} \Bigr) \Bigr]^2 \leq 1.
 \label{WPduality-ours}
\end{eqnarray}
Note that $\GRAB$ is proportional to the classical solution of field at Alice sourced by Bob while $\overline{\chi_\mathrm{B}}$ is the classical solution of field 
at Bob sourced by Alice.
The first term in Eq.~\eqref{WPduality-ours} represents the visibility of Alice and becomes one in the adiabatic limit. In this case, the second term must vanish and Bob cannot distinguish Alice's spin. It occurs since Alice cannot generate a sufficient amount of field configurations by which Bob distinguishes Alice's spin. On the other hand, if Bob can get sufficient information to distinguish Alice's spin, the visibility of Alice must vanish by emitting a large number of on-shell particles. 

 \section{Conclusions}\label{Sec:conclusion}
 Quantum mechanics has complementarity: in some situations, 
a system is in a coherent superposition of states
and, in another, interference is decohered by measurement, and the state is localized. 
In relativistic quantum theories, complementarity coexists in a  consistent way with relativistic causality. 
In this paper, we explicitly investigated the gedanken experiment  \cite{Belenchia:2018szb,Belenchia:2019gcc, Danielson:2021egj} by using
CTP formalism in relativistic quantum field theories. 
A key ingredient is the different roles of various  Green's functions. 
Causality is associated with a retarded Green's function, while quantum fluctuations or 
emission of radiation are associated with the Keldysh one. Both of them play 
similarly important roles in the analysis, as seen in Eq.~\eqref{finalW}. 
Our result in Eq.~\eqref{result-c} shows 
that an expectation value  $\expval{\sigma_x}=\Re \braket{ \Psi_\down}{\Psi_\up }_{\phi, \mathrm{B}}$, which represents
decoherence of a coherently superposed state,  is given by two independent
effects of particle emission $e^{-\Gamma_{\mathrm{A}}}$ and measurement $\delta_\epsilon(\GRAB)$.
Each of them is expressed by $G_{\mathrm{K}}$ and $G_\mathrm{R}$, respectively.
The advanced Green's function also implicitly played an important role in the calculation. 
The visibility $v$ in Eq.~\eqref{result-c} has a trade-off relation (\ref{WD-Bob}) with the distinguishability of Bob $D_\mathrm{B}$. Thus, in order for Bob to sufficiently distinguish Alice's spin, Alice must emit on-shell particles and the visibility must be lost.  


 The same logic argued in the paper can be applied to any relativistic quantum field theories, including gravity
 as far as linearized approximation is valid. 
 The only differences are  concrete forms of Green's functions and couplings between fields and matter.  
 
In our calculation, decoherence is induced by two effects $e^{-\Gamma_{\mathrm{A}}}$ and $ \delta_\epsilon (\GRAB)$, one by Alice's nonadiabaticity and the other by Bob's measurement, and the quantum nature of fields is responsible for both effects. The emission of radiation occurs even classically
but the randomness of phase evolution is associated with the quantum fluctuations of fields in the vacuum. Indeed, 
$\Gamma_{\mathrm{A}}$ is of order $\order{\hbar}$ compared to the classical action. 
On the other hand, the decoherence caused by Bob $\delta_\epsilon (\GRAB)$ reflects the fact that the field itself is in an entangled state correlated with spin, and is essentially quantum. Thus, if we can measure similar effects of 
decoherence in the gravitational case, it is evidence of the quantum nature of gravity. 
\\

\begin{acknowledgments}
The work initiated from discussions at the QUP meetings on quantum sensors and their
particle physics applications. We thank all participants in the meetings for the discussions.
We especially  thank Masahiro Hotta, Sugumi Kanno, Yasusada Nambu, Jiro Soda, Izumi Tsutsui and Kazuhiro Yamamoto for their indispensable information on the subject and fruitful discussions. 
S.I. is supported in part by the Grant-in-Aid for Scientific research, No. 18H03708 and No. 16H06490.
Y.H. is supported in part by the Grant-in-Aid for Scientific research, No. 21H01084.  
\end{acknowledgments}

\appendix

\section{Particle production}\label{sec:particle_production}
We will show the number of created particles $\expval{n}$ produced by Alice's protocol $\lambda_\mathrm{A}(t)$ is equal to $\Gamma_\mathrm{A}/2$ in Eq.~\eqref{eq:GammaA}.
In general, the number of created particles in the presence of an external field $J$ is given as~\cite{Gelis:2006yv,Gelis:2006cr}\footnote{The expression might look slightly different from that in Refs.~\cite{Gelis:2006yv,Gelis:2006cr}. There is $(-)$ on the right-hand side of Eq.~\eqref{eq:multiplicity}, which comes from our convention $J_2=-J^2$.
Therefore, both expressions are equivalent.}
\begin{equation}
  \begin{split}
    \expval{n}&=-\int\dd[4]x\int\dd[4]x' ZG_{12}(x,x')\\
    &\frac{(-\partial_x^2+m^2)}{Z}
    \frac{(-\partial_{x'}^2+m^2)}{Z}
     \qty[\frac{\delta  \ri  W}{\delta  J^1(x)}  \frac{\delta \ri W}{\delta  J^2(x')}
     +  \frac{\delta  \ri W}{\delta  J^1(x)\delta  J^2(x')} ]_{J_1=J_2=J},
  \end{split}
  \label{eq:multiplicity}
\end{equation}
where $Z$ represents the wave function renormalization factor. 
The term following $G_{12}(x,x')$
is the probability producing two particles at $x$ and $x^\prime$ due to the Lehmann-Symanzik-Zimmermann formula. The integration with the Wightman Green's function $G_{12}$ is nothing but the integration over all the final states with an appropriate normalization of wave functions, as is clear in the momentum space. 
For free theory, the formula is drastically simplified. First $Z=1$. 
When it is coupled with an external field,
noting from Eq.~\eqref{eq:generating_functional}, we have
\begin{equation}
    (-\partial_x^2+m^2)\frac{\delta  \ri  W}{\delta  J^i(x)}
    =-(-\partial_x^2+m^2)\int\dd[4]{x}G_{ij}(x,y)J^j(y)
    = \ri J_i(x),
\end{equation}
and
\begin{equation}
    (-\partial_x^2+m^2)\frac{\delta  \ri W}{\delta  J^1(x)\delta  J^2(x')} 
    =-(-\partial_x^2+m^2) G_{12}(x,x')=0,
\end{equation}
we obtain
\begin{equation}
    \begin{split}
  \expval{n} &= \int\dd[4]x\int\dd[4]x' J(x)G_{12}(x,x') J(x')\\
  &= 
   \int\dd[4]x\int\dd[4]x' J(x)G_{\mathrm{K}}(x,x') J(x').
   \end{split}
\end{equation}
In the second line, we symmetrized the propagator and used the relation $(G_{12}(x,x')+G_{21}(x,x'))/2=G_\mathrm{K}(x,x')$.
Therefore, setting $J(x)= \lambda_{\mathrm{A}}(t) \delta^{(3)}(\vb*{x}-\vb*{x}_{\mathrm{A}})$,
we find $\expval{n}=\Gamma_\mathrm{A}/2$.
We note that this relation is a special case of no self-interaction of fields.
If the interaction exists, the factor of the decoherence and the number of created particles will not be proportional.

\section{Effect of the field on the distinguishability} \label{subsec-fieldeffect}
The inequality $D_\mathrm{B} \leq D_{\mathrm{B}, \phi}$ is generated when the emission  of radiation of the field $\phi$ makes Bob's measurement obscure. 
The density operators $\rho_{\mathrm{B},\phi}^{\updown}$ describe pure states $\ket{\Psi_{\updown}}$, respectively.
Since they are not orthogonal, we can write 
\begin{eqnarray}
 \ket{\Psi_{\down}} = e^{\ri \varphi} \cos \theta \ket{\Psi_{\up}} + \sin \theta \ket*{\tilde{\Psi}_{\down}},
\end{eqnarray}
where $\braket* {\Psi_{\up}} {\tilde{\Psi}_{\down}} =0$.
Then 
\begin{eqnarray}
 \rho_{\mathrm{B},\phi}^{\up} - \rho_{\mathrm{B},\phi}^{\down} &= 
 \begin{pmatrix}
  \ket{\Psi_{\up}}   &   \ket{\tilde{\Psi}_{\down}}
 \end{pmatrix}
 \begin{pmatrix}
\sin^2 \theta  & - e^{\ri \varphi} \cos \theta \sin \theta\\
 - e^{-\ri \varphi} \cos \theta \sin \theta     & -\sin^2 \theta
 \end{pmatrix}
\begin{pmatrix}
   \bra{\Psi_{\up}}     \\
       \bra{\tilde{\Psi}_{\down}}
\end{pmatrix}.
\end{eqnarray} 
Eigenvalues of the matrix are given by $\pm \sin \theta$. Thus, $D_{\mathrm{B},\phi}= |\sin \theta|$. As visibility is given by $v^2=\cos^2 \theta$, we have an equality 
$v^2 + D_{\mathrm{B},\phi}^2=1.$ 
Writing the eigenstates as $\ket{\alpha}$ and $\ket{\beta}$, the distinguishability of Bob is given by
taking the partial trace over the field before taking the absolute value as
\begin{eqnarray}
 D_\mathrm{B} &=|\sin \theta|  \times \frac{1}{2}  \tr_\mathrm{B} \left|  \left( \tr_\phi\ket{\alpha}\bra{\alpha} - \tr_\phi\ket{\beta}\bra{\beta} \right) \right|
 \le |\sin \theta|.
\end{eqnarray}
The equality holds only if $\tr_\mathrm{B} \left[ (\tr_\phi\ket{\alpha}\bra{\alpha})( \tr_\phi\ket{\beta}\bra{\beta} ) \right]=0$.

\bibliographystyle{apsrev4-1}
\bibliography{gravity}

\begin{thebibliography}{29}%
\makeatletter
\providecommand \@ifxundefined [1]{%
 \@ifx{#1\undefined}
}%
\providecommand \@ifnum [1]{%
 \ifnum #1\expandafter \@firstoftwo
 \else \expandafter \@secondoftwo
 \fi
}%
\providecommand \@ifx [1]{%
 \ifx #1\expandafter \@firstoftwo
 \else \expandafter \@secondoftwo
 \fi
}%
\providecommand \natexlab [1]{#1}%
\providecommand \enquote  [1]{``#1''}%
\providecommand \bibnamefont  [1]{#1}%
\providecommand \bibfnamefont [1]{#1}%
\providecommand \citenamefont [1]{#1}%
\providecommand \href@noop [0]{\@secondoftwo}%
\providecommand \href [0]{\begingroup \@sanitize@url \@href}%
\providecommand \@href[1]{\@@startlink{#1}\@@href}%
\providecommand \@@href[1]{\endgroup#1\@@endlink}%
\providecommand \@sanitize@url [0]{\catcode `\\12\catcode `\$12\catcode
  `\&12\catcode `\#12\catcode `\^12\catcode `\_12\catcode `\%12\relax}%
\providecommand \@@startlink[1]{}%
\providecommand \@@endlink[0]{}%
\providecommand \url  [0]{\begingroup\@sanitize@url \@url }%
\providecommand \@url [1]{\endgroup\@href {#1}{\urlprefix }}%
\providecommand \urlprefix  [0]{URL }%
\providecommand \Eprint [0]{\href }%
\providecommand \doibase [0]{http://dx.doi.org/}%
\providecommand \selectlanguage [0]{\@gobble}%
\providecommand \bibinfo  [0]{\@secondoftwo}%
\providecommand \bibfield  [0]{\@secondoftwo}%
\providecommand \translation [1]{[#1]}%
\providecommand \BibitemOpen [0]{}%
\providecommand \bibitemStop [0]{}%
\providecommand \bibitemNoStop [0]{.\EOS\space}%
\providecommand \EOS [0]{\spacefactor3000\relax}%
\providecommand \BibitemShut  [1]{\csname bibitem#1\endcsname}%
\let\auto@bib@innerbib\@empty
\bibitem [{\citenamefont {Abbott}\ \emph {et~al.}(2016)\citenamefont {Abbott}
  \emph {et~al.}}]{LIGOScientific:2016aoc}%
  \BibitemOpen
  \bibfield  {author} {\bibinfo {author} {\bibfnamefont {B.~P.}\ \bibnamefont
  {Abbott}} \emph {et~al.} (\bibinfo {collaboration} {LIGO Scientific,
  Virgo}),\ }\href {\doibase 10.1103/PhysRevLett.116.061102} {\bibfield
  {journal} {\bibinfo  {journal} {Phys. Rev. Lett.}\ }\textbf {\bibinfo
  {volume} {116}},\ \bibinfo {pages} {061102} (\bibinfo {year} {2016})},\
  \Eprint {http://arxiv.org/abs/1602.03837} {arXiv:1602.03837 [gr-qc]}
  \BibitemShut {NoStop}%
\bibitem [{\citenamefont {Jacobson}(1995)}]{Jacobson:1995ab}%
  \BibitemOpen
  \bibfield  {author} {\bibinfo {author} {\bibfnamefont {T.}~\bibnamefont
  {Jacobson}},\ }\href {\doibase 10.1103/PhysRevLett.75.1260} {\bibfield
  {journal} {\bibinfo  {journal} {Phys. Rev. Lett.}\ }\textbf {\bibinfo
  {volume} {75}},\ \bibinfo {pages} {1260} (\bibinfo {year} {1995})},\ \Eprint
  {http://arxiv.org/abs/gr-qc/9504004} {arXiv:gr-qc/9504004} \BibitemShut
  {NoStop}%
\bibitem [{\citenamefont {Verlinde}(2011)}]{Verlinde:2010hp}%
  \BibitemOpen
  \bibfield  {author} {\bibinfo {author} {\bibfnamefont {E.~P.}\ \bibnamefont
  {Verlinde}},\ }\href {\doibase 10.1007/JHEP04(2011)029} {\bibfield  {journal}
  {\bibinfo  {journal} {JHEP}\ }\textbf {\bibinfo {volume} {04}},\ \bibinfo
  {pages} {029} (\bibinfo {year} {2011})},\ \Eprint
  {http://arxiv.org/abs/1001.0785} {arXiv:1001.0785 [hep-th]} \BibitemShut
  {NoStop}%
\bibitem [{\citenamefont {Howl}\ \emph {et~al.}(2018)\citenamefont {Howl},
  \citenamefont {Hackerm\"uller}, \citenamefont {Bruschi},\ and\ \citenamefont
  {Fuentes}}]{Howl:2016ryt}%
  \BibitemOpen
  \bibfield  {author} {\bibinfo {author} {\bibfnamefont {R.}~\bibnamefont
  {Howl}}, \bibinfo {author} {\bibfnamefont {L.}~\bibnamefont
  {Hackerm\"uller}}, \bibinfo {author} {\bibfnamefont {D.~E.}\ \bibnamefont
  {Bruschi}}, \ and\ \bibinfo {author} {\bibfnamefont {I.}~\bibnamefont
  {Fuentes}},\ }\href {\doibase 10.1080/23746149.2017.1383184} {\bibfield
  {journal} {\bibinfo  {journal} {Adv. Phys. X}\ }\textbf {\bibinfo {volume}
  {3}},\ \bibinfo {pages} {1383184} (\bibinfo {year} {2018})},\ \Eprint
  {http://arxiv.org/abs/1607.06666} {arXiv:1607.06666 [quant-ph]} \BibitemShut
  {NoStop}%
\bibitem [{\citenamefont {Bose}\ \emph {et~al.}(2017)\citenamefont {Bose},
  \citenamefont {Mazumdar}, \citenamefont {Morley}, \citenamefont {Ulbricht},
  \citenamefont {Toro\v{s}}, \citenamefont {Paternostro}, \citenamefont
  {Geraci}, \citenamefont {Barker}, \citenamefont {Kim},\ and\ \citenamefont
  {Milburn}}]{Bose:2017nin}%
  \BibitemOpen
  \bibfield  {author} {\bibinfo {author} {\bibfnamefont {S.}~\bibnamefont
  {Bose}}, \bibinfo {author} {\bibfnamefont {A.}~\bibnamefont {Mazumdar}},
  \bibinfo {author} {\bibfnamefont {G.~W.}\ \bibnamefont {Morley}}, \bibinfo
  {author} {\bibfnamefont {H.}~\bibnamefont {Ulbricht}}, \bibinfo {author}
  {\bibfnamefont {M.}~\bibnamefont {Toro\v{s}}}, \bibinfo {author}
  {\bibfnamefont {M.}~\bibnamefont {Paternostro}}, \bibinfo {author}
  {\bibfnamefont {A.}~\bibnamefont {Geraci}}, \bibinfo {author} {\bibfnamefont
  {P.}~\bibnamefont {Barker}}, \bibinfo {author} {\bibfnamefont {M.~S.}\
  \bibnamefont {Kim}}, \ and\ \bibinfo {author} {\bibfnamefont
  {G.}~\bibnamefont {Milburn}},\ }\href {\doibase
  10.1103/PhysRevLett.119.240401} {\bibfield  {journal} {\bibinfo  {journal}
  {Phys. Rev. Lett.}\ }\textbf {\bibinfo {volume} {119}},\ \bibinfo {pages}
  {240401} (\bibinfo {year} {2017})},\ \Eprint
  {http://arxiv.org/abs/1707.06050} {arXiv:1707.06050 [quant-ph]} \BibitemShut
  {NoStop}%
\bibitem [{\citenamefont {Marletto}\ and\ \citenamefont
  {Vedral}(2017)}]{Marletto:2017kzi}%
  \BibitemOpen
  \bibfield  {author} {\bibinfo {author} {\bibfnamefont {C.}~\bibnamefont
  {Marletto}}\ and\ \bibinfo {author} {\bibfnamefont {V.}~\bibnamefont
  {Vedral}},\ }\href {\doibase 10.1103/PhysRevLett.119.240402} {\bibfield
  {journal} {\bibinfo  {journal} {Phys. Rev. Lett.}\ }\textbf {\bibinfo
  {volume} {119}},\ \bibinfo {pages} {240402} (\bibinfo {year} {2017})},\
  \Eprint {http://arxiv.org/abs/1707.06036} {arXiv:1707.06036 [quant-ph]}
  \BibitemShut {NoStop}%
\bibitem [{\citenamefont {Miki}\ \emph {et~al.}(2021)\citenamefont {Miki},
  \citenamefont {Matsumura},\ and\ \citenamefont {Yamamoto}}]{Miki:2020hvg}%
  \BibitemOpen
  \bibfield  {author} {\bibinfo {author} {\bibfnamefont {D.}~\bibnamefont
  {Miki}}, \bibinfo {author} {\bibfnamefont {A.}~\bibnamefont {Matsumura}}, \
  and\ \bibinfo {author} {\bibfnamefont {K.}~\bibnamefont {Yamamoto}},\ }\href
  {\doibase 10.1103/PhysRevD.103.026017} {\bibfield  {journal} {\bibinfo
  {journal} {Phys. Rev. D}\ }\textbf {\bibinfo {volume} {103}},\ \bibinfo
  {pages} {026017} (\bibinfo {year} {2021})},\ \Eprint
  {http://arxiv.org/abs/2010.05159} {arXiv:2010.05159 [gr-qc]} \BibitemShut
  {NoStop}%
\bibitem [{\citenamefont {Matsumura}\ and\ \citenamefont
  {Yamamoto}(2020)}]{Matsumura:2020law}%
  \BibitemOpen
  \bibfield  {author} {\bibinfo {author} {\bibfnamefont {A.}~\bibnamefont
  {Matsumura}}\ and\ \bibinfo {author} {\bibfnamefont {K.}~\bibnamefont
  {Yamamoto}},\ }\href {\doibase 10.1103/PhysRevD.102.106021} {\bibfield
  {journal} {\bibinfo  {journal} {Phys. Rev. D}\ }\textbf {\bibinfo {volume}
  {102}},\ \bibinfo {pages} {106021} (\bibinfo {year} {2020})},\ \Eprint
  {http://arxiv.org/abs/2010.05161} {arXiv:2010.05161 [gr-qc]} \BibitemShut
  {NoStop}%
\bibitem [{\citenamefont {Marshman}\ \emph {et~al.}(2020)\citenamefont
  {Marshman}, \citenamefont {Mazumdar},\ and\ \citenamefont
  {Bose}}]{Marshman:2019sne}%
  \BibitemOpen
  \bibfield  {author} {\bibinfo {author} {\bibfnamefont {R.~J.}\ \bibnamefont
  {Marshman}}, \bibinfo {author} {\bibfnamefont {A.}~\bibnamefont {Mazumdar}},
  \ and\ \bibinfo {author} {\bibfnamefont {S.}~\bibnamefont {Bose}},\ }\href
  {\doibase 10.1103/PhysRevA.101.052110} {\bibfield  {journal} {\bibinfo
  {journal} {Phys. Rev. A}\ }\textbf {\bibinfo {volume} {101}},\ \bibinfo
  {pages} {052110} (\bibinfo {year} {2020})},\ \Eprint
  {http://arxiv.org/abs/1907.01568} {arXiv:1907.01568 [quant-ph]} \BibitemShut
  {NoStop}%
\bibitem [{\citenamefont {Christodoulou}\ and\ \citenamefont
  {Rovelli}(2019)}]{Christodoulou:2018cmk}%
  \BibitemOpen
  \bibfield  {author} {\bibinfo {author} {\bibfnamefont {M.}~\bibnamefont
  {Christodoulou}}\ and\ \bibinfo {author} {\bibfnamefont {C.}~\bibnamefont
  {Rovelli}},\ }\href {\doibase 10.1016/j.physletb.2019.03.015} {\bibfield
  {journal} {\bibinfo  {journal} {Phys. Lett. B}\ }\textbf {\bibinfo {volume}
  {792}},\ \bibinfo {pages} {64} (\bibinfo {year} {2019})},\ \Eprint
  {http://arxiv.org/abs/1808.05842} {arXiv:1808.05842 [gr-qc]} \BibitemShut
  {NoStop}%
\bibitem [{\citenamefont {Bose}\ \emph {et~al.}(2022)\citenamefont {Bose},
  \citenamefont {Mazumdar}, \citenamefont {Schut},\ and\ \citenamefont
  {Toro\v{s}}}]{Bose:2022uxe}%
  \BibitemOpen
  \bibfield  {author} {\bibinfo {author} {\bibfnamefont {S.}~\bibnamefont
  {Bose}}, \bibinfo {author} {\bibfnamefont {A.}~\bibnamefont {Mazumdar}},
  \bibinfo {author} {\bibfnamefont {M.}~\bibnamefont {Schut}}, \ and\ \bibinfo
  {author} {\bibfnamefont {M.}~\bibnamefont {Toro\v{s}}},\ }\href {\doibase
  10.1103/PhysRevD.105.106028} {\bibfield  {journal} {\bibinfo  {journal}
  {Phys. Rev. D}\ }\textbf {\bibinfo {volume} {105}},\ \bibinfo {pages}
  {106028} (\bibinfo {year} {2022})},\ \Eprint
  {http://arxiv.org/abs/2201.03583} {arXiv:2201.03583 [gr-qc]} \BibitemShut
  {NoStop}%
\bibitem [{\citenamefont {Christodoulou}\ \emph {et~al.}(2022)\citenamefont
  {Christodoulou}, \citenamefont {Di~Biagio}, \citenamefont {Aspelmeyer},
  \citenamefont {Brukner}, \citenamefont {Rovelli},\ and\ \citenamefont
  {Howl}}]{Christodoulou:2022vte}%
  \BibitemOpen
  \bibfield  {author} {\bibinfo {author} {\bibfnamefont {M.}~\bibnamefont
  {Christodoulou}}, \bibinfo {author} {\bibfnamefont {A.}~\bibnamefont
  {Di~Biagio}}, \bibinfo {author} {\bibfnamefont {M.}~\bibnamefont
  {Aspelmeyer}}, \bibinfo {author} {\bibfnamefont {v.}~\bibnamefont {Brukner}},
  \bibinfo {author} {\bibfnamefont {C.}~\bibnamefont {Rovelli}}, \ and\
  \bibinfo {author} {\bibfnamefont {R.}~\bibnamefont {Howl}},\ }\href@noop {}
  {\  (\bibinfo {year} {2022})},\ \Eprint {http://arxiv.org/abs/2202.03368}
  {arXiv:2202.03368 [quant-ph]} \BibitemShut {NoStop}%
\bibitem [{\citenamefont {Sugiyama}\ \emph
  {et~al.}(2022{\natexlab{a}})\citenamefont {Sugiyama}, \citenamefont
  {Matsumura},\ and\ \citenamefont {Yamamoto}}]{Sugiyama:2022ixw}%
  \BibitemOpen
  \bibfield  {author} {\bibinfo {author} {\bibfnamefont {Y.}~\bibnamefont
  {Sugiyama}}, \bibinfo {author} {\bibfnamefont {A.}~\bibnamefont {Matsumura}},
  \ and\ \bibinfo {author} {\bibfnamefont {K.}~\bibnamefont {Yamamoto}},\
  }\href@noop {} {\  (\bibinfo {year} {2022}{\natexlab{a}})},\ \Eprint
  {http://arxiv.org/abs/2203.09011} {arXiv:2203.09011 [quant-ph]} \BibitemShut
  {NoStop}%
\bibitem [{\citenamefont {Belenchia}\ \emph {et~al.}(2018)\citenamefont
  {Belenchia}, \citenamefont {Wald}, \citenamefont {Giacomini}, \citenamefont
  {Castro-Ruiz}, \citenamefont {Brukner},\ and\ \citenamefont
  {Aspelmeyer}}]{Belenchia:2018szb}%
  \BibitemOpen
  \bibfield  {author} {\bibinfo {author} {\bibfnamefont {A.}~\bibnamefont
  {Belenchia}}, \bibinfo {author} {\bibfnamefont {R.~M.}\ \bibnamefont {Wald}},
  \bibinfo {author} {\bibfnamefont {F.}~\bibnamefont {Giacomini}}, \bibinfo
  {author} {\bibfnamefont {E.}~\bibnamefont {Castro-Ruiz}}, \bibinfo {author}
  {\bibfnamefont {v.}~\bibnamefont {Brukner}}, \ and\ \bibinfo {author}
  {\bibfnamefont {M.}~\bibnamefont {Aspelmeyer}},\ }\href {\doibase
  10.1103/PhysRevD.98.126009} {\bibfield  {journal} {\bibinfo  {journal} {Phys.
  Rev. D}\ }\textbf {\bibinfo {volume} {98}},\ \bibinfo {pages} {126009}
  (\bibinfo {year} {2018})},\ \Eprint {http://arxiv.org/abs/1807.07015}
  {arXiv:1807.07015 [quant-ph]} \BibitemShut {NoStop}%
\bibitem [{\citenamefont {Belenchia}\ \emph {et~al.}(2019)\citenamefont
  {Belenchia}, \citenamefont {Wald}, \citenamefont {Giacomini}, \citenamefont
  {Castro-Ruiz}, \citenamefont {Brukner},\ and\ \citenamefont
  {Aspelmeyer}}]{Belenchia:2019gcc}%
  \BibitemOpen
  \bibfield  {author} {\bibinfo {author} {\bibfnamefont {A.}~\bibnamefont
  {Belenchia}}, \bibinfo {author} {\bibfnamefont {R.~M.}\ \bibnamefont {Wald}},
  \bibinfo {author} {\bibfnamefont {F.}~\bibnamefont {Giacomini}}, \bibinfo
  {author} {\bibfnamefont {E.}~\bibnamefont {Castro-Ruiz}}, \bibinfo {author}
  {\bibfnamefont {v.}~\bibnamefont {Brukner}}, \ and\ \bibinfo {author}
  {\bibfnamefont {M.}~\bibnamefont {Aspelmeyer}},\ }\href {\doibase
  10.1142/S0218271819430016} {\bibfield  {journal} {\bibinfo  {journal} {Int.
  J. Mod. Phys. D}\ }\textbf {\bibinfo {volume} {28}},\ \bibinfo {pages}
  {1943001} (\bibinfo {year} {2019})},\ \Eprint
  {http://arxiv.org/abs/1905.04496} {arXiv:1905.04496 [quant-ph]} \BibitemShut
  {NoStop}%
\bibitem [{\citenamefont {Danielson}\ \emph {et~al.}(2022)\citenamefont
  {Danielson}, \citenamefont {Satishchandran},\ and\ \citenamefont
  {Wald}}]{Danielson:2021egj}%
  \BibitemOpen
  \bibfield  {author} {\bibinfo {author} {\bibfnamefont {D.~L.}\ \bibnamefont
  {Danielson}}, \bibinfo {author} {\bibfnamefont {G.}~\bibnamefont
  {Satishchandran}}, \ and\ \bibinfo {author} {\bibfnamefont {R.~M.}\
  \bibnamefont {Wald}},\ }\href {\doibase 10.1103/PhysRevD.105.086001}
  {\bibfield  {journal} {\bibinfo  {journal} {Phys. Rev. D}\ }\textbf {\bibinfo
  {volume} {105}},\ \bibinfo {pages} {086001} (\bibinfo {year} {2022})},\
  \Eprint {http://arxiv.org/abs/2112.10798} {arXiv:2112.10798 [quant-ph]}
  \BibitemShut {NoStop}%
\bibitem [{\citenamefont {Mari}\ \emph {et~al.}(2016)\citenamefont {Mari},
  \citenamefont {De~Palma},\ and\ \citenamefont {Giovannetti}}]{Mari:2015qva}%
  \BibitemOpen
  \bibfield  {author} {\bibinfo {author} {\bibfnamefont {A.}~\bibnamefont
  {Mari}}, \bibinfo {author} {\bibfnamefont {G.}~\bibnamefont {De~Palma}}, \
  and\ \bibinfo {author} {\bibfnamefont {V.}~\bibnamefont {Giovannetti}},\
  }\href {\doibase 10.1038/srep22777} {\bibfield  {journal} {\bibinfo
  {journal} {Sci. Rep.}\ }\textbf {\bibinfo {volume} {6}},\ \bibinfo {pages}
  {22777} (\bibinfo {year} {2016})},\ \Eprint {http://arxiv.org/abs/1509.02408}
  {arXiv:1509.02408 [quant-ph]} \BibitemShut {NoStop}%
\bibitem [{\citenamefont {Davies}(1977)}]{Davies}%
  \BibitemOpen
  \bibfield  {author} {\bibinfo {author} {\bibfnamefont {P.}~\bibnamefont
  {Davies}},\ }\href@noop {} {\emph {\bibinfo {title} {{The Physics of Time
  Asymmetry}}}}\ (\bibinfo  {publisher} {University of California Press},\
  \bibinfo {year} {1977})\BibitemShut {NoStop}%
\bibitem [{\citenamefont {Keldysh}(1964)}]{Keldysh:1964ud}%
  \BibitemOpen
  \bibfield  {author} {\bibinfo {author} {\bibfnamefont {L.~V.}\ \bibnamefont
  {Keldysh}},\ }\href@noop {} {\bibfield  {journal} {\bibinfo  {journal} {Zh.
  Eksp. Teor. Fiz.}\ }\textbf {\bibinfo {volume} {47}},\ \bibinfo {pages}
  {1515} (\bibinfo {year} {1964})}\BibitemShut {NoStop}%
\bibitem [{\citenamefont {Jaeger}\ \emph {et~al.}(1995)\citenamefont {Jaeger},
  \citenamefont {Shimony},\ and\ \citenamefont {Vaidman}}]{PhysRevA.51.54}%
  \BibitemOpen
  \bibfield  {author} {\bibinfo {author} {\bibfnamefont {G.}~\bibnamefont
  {Jaeger}}, \bibinfo {author} {\bibfnamefont {A.}~\bibnamefont {Shimony}}, \
  and\ \bibinfo {author} {\bibfnamefont {L.}~\bibnamefont {Vaidman}},\ }\href
  {\doibase 10.1103/PhysRevA.51.54} {\bibfield  {journal} {\bibinfo  {journal}
  {Phys. Rev. A}\ }\textbf {\bibinfo {volume} {51}},\ \bibinfo {pages} {54}
  (\bibinfo {year} {1995})}\BibitemShut {NoStop}%
\bibitem [{\citenamefont {Englert}(1996)}]{PhysRevLett.77.2154}%
  \BibitemOpen
  \bibfield  {author} {\bibinfo {author} {\bibfnamefont {B.-G.}\ \bibnamefont
  {Englert}},\ }\href {\doibase 10.1103/PhysRevLett.77.2154} {\bibfield
  {journal} {\bibinfo  {journal} {Phys. Rev. Lett.}\ }\textbf {\bibinfo
  {volume} {77}},\ \bibinfo {pages} {2154} (\bibinfo {year}
  {1996})}\BibitemShut {NoStop}%
\bibitem [{\citenamefont {Sugiyama}\ \emph
  {et~al.}(2022{\natexlab{b}})\citenamefont {Sugiyama}, \citenamefont
  {Matsumura},\ and\ \citenamefont {Yamamoto}}]{Sugiyama:2022wcd}%
  \BibitemOpen
  \bibfield  {author} {\bibinfo {author} {\bibfnamefont {Y.}~\bibnamefont
  {Sugiyama}}, \bibinfo {author} {\bibfnamefont {A.}~\bibnamefont {Matsumura}},
  \ and\ \bibinfo {author} {\bibfnamefont {K.}~\bibnamefont {Yamamoto}},\
  }\href@noop {} {\  (\bibinfo {year} {2022}{\natexlab{b}})},\ \Eprint
  {http://arxiv.org/abs/2206.02506} {arXiv:2206.02506 [quant-ph]} \BibitemShut
  {NoStop}%
\bibitem [{\citenamefont {Kanno}\ \emph {et~al.}(2021)\citenamefont {Kanno},
  \citenamefont {Soda},\ and\ \citenamefont {Tokuda}}]{Kanno:2020usf}%
  \BibitemOpen
  \bibfield  {author} {\bibinfo {author} {\bibfnamefont {S.}~\bibnamefont
  {Kanno}}, \bibinfo {author} {\bibfnamefont {J.}~\bibnamefont {Soda}}, \ and\
  \bibinfo {author} {\bibfnamefont {J.}~\bibnamefont {Tokuda}},\ }\href
  {\doibase 10.1103/PhysRevD.103.044017} {\bibfield  {journal} {\bibinfo
  {journal} {Phys. Rev. D}\ }\textbf {\bibinfo {volume} {103}},\ \bibinfo
  {pages} {044017} (\bibinfo {year} {2021})},\ \Eprint
  {http://arxiv.org/abs/2007.09838} {arXiv:2007.09838 [hep-th]} \BibitemShut
  {NoStop}%
\bibitem [{\citenamefont {Schwinger}(1961)}]{Schwinger:1960qe}%
  \BibitemOpen
  \bibfield  {author} {\bibinfo {author} {\bibfnamefont {J.~S.}\ \bibnamefont
  {Schwinger}},\ }\href {\doibase 10.1063/1.1703727} {\bibfield  {journal}
  {\bibinfo  {journal} {J. Math. Phys.}\ }\textbf {\bibinfo {volume} {2}},\
  \bibinfo {pages} {407} (\bibinfo {year} {1961})}\BibitemShut {NoStop}%
\bibitem [{\citenamefont {Rammer}(2007)}]{Rammer:2007zz}%
  \BibitemOpen
  \bibfield  {author} {\bibinfo {author} {\bibfnamefont {J.}~\bibnamefont
  {Rammer}},\ }\href@noop {} {\emph {\bibinfo {title} {{Quantum field theory of
  non-equilibrium states}}}}\ (\bibinfo  {publisher} {{Cambridge University
  Press}},\ \bibinfo {year} {2007})\BibitemShut {NoStop}%
\bibitem [{\citenamefont
  {Mart\'{\i}n-Mart\'{\i}nez}(2015)}]{PhysRevD.92.104019}%
  \BibitemOpen
  \bibfield  {author} {\bibinfo {author} {\bibfnamefont {E.}~\bibnamefont
  {Mart\'{\i}n-Mart\'{\i}nez}},\ }\href {\doibase 10.1103/PhysRevD.92.104019}
  {\bibfield  {journal} {\bibinfo  {journal} {Phys. Rev. D}\ }\textbf {\bibinfo
  {volume} {92}},\ \bibinfo {pages} {104019} (\bibinfo {year}
  {2015})}\BibitemShut {NoStop}%
\bibitem [{\citenamefont {Nielsen}\ and\ \citenamefont
  {Chuang}(2011)}]{10.5555/1972505}%
  \BibitemOpen
  \bibfield  {author} {\bibinfo {author} {\bibfnamefont {M.~A.}\ \bibnamefont
  {Nielsen}}\ and\ \bibinfo {author} {\bibfnamefont {I.~L.}\ \bibnamefont
  {Chuang}},\ }\href@noop {} {\emph {\bibinfo {title} {Quantum Computation and
  Quantum Information: 10th Anniversary Edition}}},\ \bibinfo {edition} {10th}\
  ed.\ (\bibinfo  {publisher} {Cambridge University Press},\ \bibinfo {address}
  {USA},\ \bibinfo {year} {2011})\BibitemShut {NoStop}%
\bibitem [{\citenamefont {Gelis}\ and\ \citenamefont
  {Venugopalan}(2006{\natexlab{a}})}]{Gelis:2006yv}%
  \BibitemOpen
  \bibfield  {author} {\bibinfo {author} {\bibfnamefont {F.}~\bibnamefont
  {Gelis}}\ and\ \bibinfo {author} {\bibfnamefont {R.}~\bibnamefont
  {Venugopalan}},\ }\href {\doibase 10.1016/j.nuclphysa.2006.07.020} {\bibfield
   {journal} {\bibinfo  {journal} {Nucl. Phys. A}\ }\textbf {\bibinfo {volume}
  {776}},\ \bibinfo {pages} {135} (\bibinfo {year} {2006}{\natexlab{a}})},\
  \Eprint {http://arxiv.org/abs/hep-ph/0601209} {arXiv:hep-ph/0601209}
  \BibitemShut {NoStop}%
\bibitem [{\citenamefont {Gelis}\ and\ \citenamefont
  {Venugopalan}(2006{\natexlab{b}})}]{Gelis:2006cr}%
  \BibitemOpen
  \bibfield  {author} {\bibinfo {author} {\bibfnamefont {F.}~\bibnamefont
  {Gelis}}\ and\ \bibinfo {author} {\bibfnamefont {R.}~\bibnamefont
  {Venugopalan}},\ }\href {\doibase 10.1016/j.nuclphysa.2006.08.015} {\bibfield
   {journal} {\bibinfo  {journal} {Nucl. Phys. A}\ }\textbf {\bibinfo {volume}
  {779}},\ \bibinfo {pages} {177} (\bibinfo {year} {2006}{\natexlab{b}})},\
  \Eprint {http://arxiv.org/abs/hep-ph/0605246} {arXiv:hep-ph/0605246}
  \BibitemShut {NoStop}%
\end{thebibliography}%
\end{document}